\documentclass[10pt, pra, aps, twocolumn, showpacs, citeautoscript, floatfix, reprint, amsmath, amssymb, notitlepage, superscriptaddress]{revtex4-2}

\usepackage{graphicx}
\usepackage{stmaryrd}
\usepackage{rotating}
\usepackage{amsmath}
\usepackage{amsfonts}
\usepackage{amssymb}
\usepackage{wasysym}
\usepackage[countmax]{subfloat}
\usepackage{dcolumn} 
\usepackage{bm} 
\usepackage{color}

\usepackage{float}
\usepackage{latexsym}

\begin{document}

\title{Quantum Cram\'{e}r-Rao bound on quantum metric as a single-observable \\ uncertainty relation}









\author{Wei Chen}

\affiliation{Department of Physics, PUC-Rio, 22451-900 Rio de Janeiro, Brazil}

\date{\rm\today}

\begin{abstract}

The Belavkin version of quantum Cram\'{e}r-Rao bound dictates that the covariance of any set of operators is bounded by a product of the derivatives of expectation values and the inverse of quantum metric. We elaborate that because quantum metric itself is the covariance of the generators of translation in the parameter space, quantum metric in any dimension is bounded by a product of itself and Berry curvature. The generator formalism further indicates that the bound is equivalent to a single-observable uncertainty relation for systems described by multiple observables, which in the two-observable case recovers the Robertson-Schr\"{o}dinger uncertainty relation. The momentum space quantum metric and spin operators of three-dimensional topological insulators under magnetic field are used to demonstrate the validity of these bounds, and their constraints on the spread of Wannier function, absorbance in 2D, and dielectric function in 3D will be elaborated. 




\end{abstract}

\maketitle

\section{Introduction}

A celebrated milestone in quantum metrology is the discovery of quantum Cram\'{e}r-Rao bound (QCRB)\cite{Helstrom67,Braunstein94,Braunstein96}. Analogous to its classical counter part\cite{Rao45,Cramer46}, the QCRB manifests in the problem of quantum parameter estimation, i.e., when one intends to extract some parameters from a set of data obtained from certain quantum measurements. The QCRB states that the precision of any unbiased estimators for the parameters is limited by the quantum Fisher information matrix (QFIM)\cite{Helstrom67,Hovelo11,Braunstein94,Braunstein96,Paris09,Liu20}, which is a generalization of the classical Fisher information in the information geometry to the quantum setting\cite{Fisher25,Amari16}, usually constructed from the density matrix of the quantum system under question. This paradigm of quantum parameter estimation manifests in all sorts of quantum measurements, and enjoys a surge of interest owing to recent advances in quantum information and quantum computation. 


In this paper, we demonstrate that an operator version of QCRB, first formulated by Belavkin\cite{Belavkin76}, results in several geometric bounds that constrain many experimental observables in condensed matter physics. The Belavkin version of QCRB dictates that the covariance of any set of Hermitian operators in a quantum system is bounded by a product of the derivatives of the expectation values and the inverse of QFIM. In fact, the usual QCRB in quantum parameter estimation can be regarded as a special case of the Belavkin version when the Hermitian operators are taken to be the unbiased estimators of the parameters\cite{Paris09,Liu20}. Particularly for pure states whose density matrices only consist of a single state, if we take the Hermitian operators to be the generators of translation in the parameter space, then their covariances give the quantum metric\cite{Provost80}, often also referred to as the fidelity susceptibility\cite{You07,Zanardi07}, a concept that quantifies the fidelity between two quantum states that are infinitely close in the parameter space. Moreover, the derivatives on the expectation values of the generators are equivalent to the Berry curvature\cite{Berry84,Xiao10}, and the QFIM reduces to four times the quantum metric in this pure state limit. As a result, the Belavkin version of QCRB manifests as a geometric bound on quantum metric and Berry curvature in any parameter space of arbitrary dimension. The coincidence of this QCRB and a well-known bound on quantum metric for 2D parameter space will be elaborated\cite{Roy14,Peotta15,Ozawa21}. Motivated by a recent surge of interest on the momentum space quantum metric of semiconductors and insulators, we will use 3D topological insulators (TIs) in a magnetic field as an example to demonstrate the validity of this self-bound on quantum metric in the momentum space treated as a 3D parameter space. Furthermore, owing to a relation between quantum metric and the spread of Wannier function\cite{Marzari97,Marzari12,Porlles23_singlet_SC_metric}, and the fact that quantum metric coincides with the optical transition matrix element\cite{Ozawa18,Ahn22}, we will elaborate that the spread of Wannier function, absorbance in 2D, and dielectric function in 3D in each spatial direction of semiconductors are all constrained by our bound on quantum metric, indicating its significant influence on optical properties of solids. 


We further elaborate that the Belavkin version of QCRB implies a variety of uncertainty relations. Because the quantum metric is the covariance of generators while Berry curvature is the commutator of the generators\cite{Provost80}, the Belavkin version of QCRB can be written entirely as a bound on the variance of a generator caused by a product of covariances and commutators of all the generators in the parameter space. The advantage of this generator scenario is that precisely what is the parameter space that the generators translate is unimportant, and thus any Hermitian operator can be regarded as a generator in some space. As a result, one is lead to a single-observable uncertainty relation that bounds the variance of each individual operator by the covariances and commutators of all the operators in the system. Note that although many uncertainty relations for the sum or product of the variances of multiple observables have been derived\cite{Robertson34,Synge71,Maccone14,Chen15,Qin16,Song16,Song17,Chen17,Dodonov18,Dodonov25}, one cannot naively subtract or divide them to obtain the constraint on individual observables, pointing to the uniqueness of our single-observable uncertainty relation. The two-operator version of our formula recovers the usual Robertson-Schr\"{o}dinger uncertainty relation\cite{Robertson29,Schrodinger30}, suggesting the interpretation of Robertson-Schr\"{o}dinger uncertainty relation in terms of the Belavkin version of QCRB. We will use the three spin operators in the aforementioned 3D TI in a magnetic field as an example of the three-operator situation, which confirm that our single-observable uncertainty relation is always satisfied in any situation.

\section{Quantum Cram\'{e}r-Rao bound on variance of operators}

\subsection{Geometric interpretation \label{sec:geometric_interpretation}}

Our main goal is to apply the Belavkin version of QCRB\cite{Belavkin76}, originally formulated for mixed states, to quantum geometry of pure states in condensed matter systems. For clarity, we first review the original mixed-state formalism, and later discuss pure states as a limit of this formalism. Consider a mixed state described by the density matrix $\rho$ defined in a $D$-dimensional parameter space ${\bf k}=(k^{1},k^{2}...k^{D})$
\begin{eqnarray}
\rho({\bf k})=\sum_{i}\lambda_{i}({\bf k})|i({\bf k})\rangle\langle i({\bf k})|,
\end{eqnarray}
where we have expanded the density matrix in terms of its eigenstates $|i({\bf k})\rangle$ and eigenvalues $\lambda_{i}({\bf k})$. Denoting the derivative with respect to a specific parameter by $\partial_{\mu}=\partial/\partial k^{\mu}$, the derivative on the density matrix defines the symmetric logarithmic derivative (SLD) $L_{\mu}$
\begin{eqnarray}
\partial_{\mu}\rho=\frac{1}{2}\left(\rho L_{\mu}+L_{\mu}\rho\right).
\label{SLD_definition}
\end{eqnarray}
from which the QFIM is introduced\cite{Helstrom67,Hovelo11,Braunstein94,Braunstein96} 
\begin{eqnarray}
F_{\mu\nu}({\bf k})=\frac{1}{2}{\rm Tr}\rho\left\{L_{\mu},L_{\nu}\right\}.
\label{QFIM_matrix_element}
\end{eqnarray}
Consider a set of Hermitian operators ${\hat O}_{a}=\left\{{\hat O}_{1},{\hat O}_{2}...\right\}$ that may or may not depend on the parameter ${\bf k}$, and we denote ${\bf C}$ with matrix element $C_{ab}=\frac{1}{2}\langle\left\{\Delta{\hat O}_{a},\Delta{\hat O}_{b}\right\}\rangle$ as the covariance matrix of the Hermitian operators. Each operator has the expectation value $\langle \hat{O}_{a} \rangle = \rm{Tr} \rho\,\hat{O}_{a}$ that depends on the parameter ${\bf k}$. The Belavkin version of QCRB states that the covariance matrix ${\bf C}$ of any set of operators satisfies the inequality\cite{Belavkin76}
\begin{eqnarray}
{\bf C}\geq {\rm Tr}({\boldsymbol\nabla}\rho\,{\hat {\bf O}})^{T}\cdot{\bf F}^{-1}\cdot {\rm Tr}({\boldsymbol\nabla}\rho\,{\hat {\bf O}})\geq 0.
\label{QCRB_operator_mixed_state}
\end{eqnarray}
All the diagonal elements of this inequality have to be positive, and hence (repeated Greek indices are summed)
\begin{eqnarray}
\langle\Delta{\hat O}_{a}^{2}\rangle\geq
 {\rm Tr}({\hat O}_{a}\partial_{\mu}\rho) F^{\mu\nu}{\rm Tr}(\partial_{\nu}\rho\,{\hat O}_{a})\geq 0,
\label{QCRB_single_operator}
\end{eqnarray}
where $F^{\mu\nu}$ is the inverse of QFIM satisfying $F_{\mu\nu}F^{\nu\sigma}=\delta_{\mu}^{\sigma}$. 


For pure states that are the focus of this work, the density matrix contains only one eigenstate $|\psi({\bf k})\rangle$. The density matrix, SLD, expectation value of the Hermitian operator, and its derivative for pure states are given by
\begin{eqnarray}
&&\rho=|\psi\rangle\langle\psi|,
\nonumber \\
&&L_{\mu}=2|\partial_{\mu}\psi\rangle\langle\psi|+2|\psi\rangle\langle\partial_{\mu}\psi|,
\nonumber \\
&&\langle{\hat O}_{a}\rangle=\langle\psi|{\hat O}_{a}|\psi\rangle,
\nonumber \\
&&{\rm Tr}(\partial_{\mu}\rho\,{\hat O}_{a})=\langle\partial_{\mu}\psi|{\hat O}_{a}|\psi\rangle+\langle\psi|{\hat O}_{a}|\partial_{\mu}\psi\rangle.
\label{pure_state_rho_dTrO}
\end{eqnarray}
The notion of quantum geometry of pure states arises from the quantum metric and Berry curvature of the state $|\psi({\bf k})\rangle$ that take the form\cite{Provost80,Berry84,Xiao10}
\begin{eqnarray}
&&g_{\mu\nu}=\frac{1}{2}\langle\partial_{\mu}\psi|\partial_{\nu}\psi\rangle
+\frac{1}{2}\langle\partial_{\nu}\psi|\partial_{\mu}\psi\rangle
-\langle\partial_{\mu}\psi|\psi\rangle\langle\psi|\partial_{\nu}\psi\rangle,
\nonumber \\
&&\Omega_{\mu\nu}=i\langle\partial_{\mu}\psi|\partial_{\nu}\psi\rangle
-i\langle\partial_{\nu}\psi|\partial_{\mu}\psi\rangle.
\label{gmunu_Omegamunu_psi_formula}
\end{eqnarray}
Inserting Eq.~(\ref{pure_state_rho_dTrO}) into Eqs.~(\ref{SLD_definition}) and (\ref{QFIM_matrix_element}), one sees that the QFIM is equal to four times the quantum metric $F_{\mu\nu}\rightarrow 4g_{\mu\nu}$ for pure states, and hence its inverse reduces to $F^{\mu\nu}\rightarrow g^{\mu\nu}/4$. Putting these pure-state results into Eq.~(\ref{QCRB_operator_mixed_state}) yields 
\begin{eqnarray}
{\bf C}\geq \frac{1}{4}{\rm Tr}({\boldsymbol\nabla}\rho\,{\hat {\bf O}})^{T}\cdot{\bf g}^{-1}\cdot {\rm Tr}({\boldsymbol\nabla}\rho\,{\hat {\bf O}})\geq 0,
\label{QCRB_operator_pure_state}
\end{eqnarray}
where ${\bf g}^{-1}$ is the inverse of the quantum metric, and we use upper indices $g^{\mu\nu}$ to denote its matrix elements that satisfy $g^{\mu\nu}g_{\nu\rho}=\delta_{\rho}^{\mu}$. The positive semidefinitiveness of Eq.~(\ref{QCRB_operator_pure_state}) implies 
\begin{eqnarray}
\langle \Delta{\hat O}_{a}^2 \rangle\geq \frac{1}{4}{\rm Tr}({\hat O}_{a}\partial_{\mu}\rho) g^{\mu\nu}{\rm Tr}(\partial_{\nu}\rho\,{\hat O}_{a})\geq 0,
\label{QCRB_operator_pure_state_single_operator}
\end{eqnarray}
valid for any operator ${\hat O}_{a}$ in the set. This inequality is the Belavkin version of QCRB applied to pure states, and it is our aim to demonstrate how it leads to nontrivial geometric bounds on the quantum metric and Berry curvature in Eq.~(\ref{gmunu_Omegamunu_psi_formula}).



The key to connect Eq.~(\ref{QCRB_operator_pure_state_single_operator}) to geometric bounds is the fact that quantum metric itself is the covariance matrix of the generators in the parameter space. This is formulated by considering the translation of the quantum state in the parameter space\cite{Provost80} 
\begin{eqnarray}
|\psi({\bf k+\delta k})\rangle=e^{-i\Lambda_{\mu}({\bf k})\delta k^{\mu}}|\psi({\bf k})\rangle,
\end{eqnarray}
where the generators $\Lambda_{\mu}({\bf k})$ satisfy
\begin{eqnarray}
&&|\partial_{\nu}\psi({\bf k})\rangle=-i\Lambda_{\nu}({\bf k})|\psi({\bf k})\rangle.
\end{eqnarray}
As a result, the quantum metric and Berry curvature correspond to the covariances and commutators of the generators 
\begin{eqnarray}
&&g_{\mu\nu}=\frac{1}{2}\langle\left\{\Lambda_{\mu},\Lambda_{\nu}\right\}\rangle
-\langle\Lambda_{\mu}\rangle\langle\Lambda_{\nu}\rangle
=\frac{1}{2}\langle\left\{\Delta\Lambda_{\mu},\Delta\Lambda_{\nu}\right\}\rangle.
\nonumber \\
&&\Omega_{\mu\nu}=i\langle\left[\Lambda_{\mu},\Lambda_{\nu}\right]\rangle
=i\langle\left[\Delta\Lambda_{\mu},\Delta\Lambda_{\nu}\right]\rangle={\rm Tr}(\partial_{\mu}\rho\,\Lambda_{\alpha}).\;\;\;\;\;\;\;\;
\label{gmunu_generator_interpretation}
\end{eqnarray}
Our observation is that, if in Eq.~(\ref{QCRB_operator_pure_state}) we choose the operators under question to be the generators ${\hat O}_{a}\rightarrow\Lambda_{\mu}$ and then apply Eq.~(\ref{gmunu_generator_interpretation}), we obtain a remarkable geometric bound on the quantum metric caused by its own inverse and Berry curvature
\begin{eqnarray}
{\bf g}\geq-\frac{1}{4}{\boldsymbol\Omega}\cdot{\bf g}^{-1}\cdot{\boldsymbol\Omega}\geq 0,
\label{selfbound_matrix_negative}
\end{eqnarray}
meaning that ${\bf g}+\frac{1}{4}{\boldsymbol\Omega}\cdot{\bf g}^{-1}\cdot{\boldsymbol\Omega}$ is positive semidefinitive. This inequality is stronger than the ${\bf g}\geq 0$ naively expected from any metric, indicating a constraint imposed by the Berry curvature in any dimension. This result also indicates that the diagonal elements of quantum metric are bounded by
\begin{eqnarray}
g_{\alpha\alpha}\geq-\frac{1}{4}\Omega_{\alpha\mu}g^{\mu\nu}\Omega_{\nu\alpha}\geq 0.
\label{selfbound_matrix_negative_diagonal} 
\end{eqnarray}
Equations (\ref{selfbound_matrix_negative}) and (\ref{selfbound_matrix_negative_diagonal}) are the main results of this work, derived by incorporating the generator formalism into the Belavkin version of QCRB, that are applicable to parameter spaces of any dimension. We proceed to invoke the adjugate matrix ${\rm adj}({\bf g})$ defined as the transpose of the cofactor matrix, which satisfies ${\bf g}\,{\rm adj}({\bf g})={\bf I}\det{\bf g}$. As a result, the inverse matrix of ${\bf g}$ can be written in terms of the adjugate matrix as
\begin{eqnarray}
{\bf g}^{-1}=\frac{1}{\det\bf{g}}{\rm adj}({\bf g}).
\label{ginverse_adjgdetg}
\end{eqnarray}
Thus one may also write Eqs.~(\ref{selfbound_matrix_negative}) as 
\begin{eqnarray}
&&4\det\bf{g}\times{\bf g}\geq-{\boldsymbol\Omega}\cdot{\rm adj}({\bf g})\cdot{\boldsymbol\Omega},
\label{QCRB_gmunu_adjugate_mat}
\end{eqnarray}
and Eq.~(\ref{selfbound_matrix_negative_diagonal}) as
\begin{eqnarray}
&&4\det{\bf g}\times g_{\alpha\alpha}\geq-\Omega_{\alpha\mu}{\rm adj}({\bf g})^{\mu\nu}\Omega_{\nu\alpha},
\label{QCRB_gmunu_adjugate_mat_diagonal}
\end{eqnarray}
in which both sides are polynomials that only contain lower indices $g_{\mu\nu}$ and $\Omega_{\mu\nu}$.



Explicitly in 2D parameter space $\mu=\left\{x,y\right\}$, Eq.~(\ref{selfbound_matrix_negative_diagonal}) renders the inequality 
\begin{eqnarray}
&&g_{xx}\geq\frac{1}{4}\Omega_{xy}^{2}g^{yy}=\frac{\Omega_{xy}^{2}g_{xx}}{4(g_{xx}g_{yy}-g_{xy}^{2})}\geq 0,
\nonumber \\
&&g_{yy}\geq\frac{1}{4}\Omega_{xy}^{2}g^{xx}=\frac{\Omega_{xy}^{2}g_{yy}}{4(g_{xx}g_{yy}-g_{xy}^{2})}\geq 0.
\label{QCRB_pure_state_2D}
\end{eqnarray}
In 3D parameter space $\mu=\left\{x,y,z\right\}$, Eq.~(\ref{selfbound_matrix_negative_diagonal}) gives
\begin{eqnarray}
&&g_{xx}\geq\frac{1}{4}\left\{\Omega_{xy}^{2}g^{yy}-2\Omega_{zx}\Omega_{xy}g^{yz}+\Omega_{zx}^{2}g^{zz}\right\}\geq 0,
\nonumber \\
&&g_{yy}\geq\frac{1}{4}\left\{\Omega_{xy}^{2}g^{xx}-2\Omega_{xy}\Omega_{yz}g^{xz}+\Omega_{yz}^{2}g^{zz}\right\}\geq 0,
\nonumber \\
&&g_{zz}\geq\frac{1}{4}\left\{\Omega_{zx}^{2}g^{xx}-2\Omega_{yz}\Omega_{zx}g^{xy}+\Omega_{yz}^{2}g^{yy}\right\}\geq 0,\;\;\;\;\;
\label{QCRB_pure_state_3D}
\end{eqnarray}
whose expressions in terms of lower indices $g_{\mu\nu}$ is given in Appendix \ref{apx:index_explicity_inequalities_3D}. The significance of these inequalities is that they constrain each individual diagonal element $g_{\mu\mu}$ of the quantum metric, signifying that the infinitesimal distance in each direction of the $D$-dimensional parameter space is bounded by a specific condition due to the parameter dependence of $|\psi({\bf k})\rangle$. 


Particularly for the 2D case, after canceling out $g_{\mu\mu}$, either one of the two inequalities in Eq.~(\ref{QCRB_pure_state_2D}) yields
\begin{eqnarray}
g_{xx} g_{yy} \geq g_{xy}^2 + \frac{1}{4} \Omega_{xy}^2, 
\label{gmunu_omegamunu_bound_2D}
\end{eqnarray}
which is a bound on the 2D quantum metric that has been derived previously by other means\cite{Roy14,Peotta15,Ozawa21}, and here we demonstrate that it can also be derived from our QCRB formalism applied to 2D parameter spaces. It should also be emphasized that Eq.~(\ref{gmunu_omegamunu_bound_2D}) is in fact the 2D version of the more generalized Robertson inequality\cite{Robertson34}
\begin{eqnarray}
\det g_{\mu\nu}\geq\det\frac{\Omega_{\mu\nu}}{2},
\label{Robertson_inequality_gmunu_Omega_munu}
\end{eqnarray}
that is useful in even dimensions (in odd dimensions $\det(\Omega_{\mu\nu}/2)=0$, yielding trivially $\det g_{\mu\nu}\geq 0$). In contrast, our inequalities in Eqs.~(\ref{selfbound_matrix_negative}) and (\ref{selfbound_matrix_negative_diagonal}) based on QCRB is applicable to any dimensions. 



\subsection{Uncertainty interpretation \label{sec:uncertainty_interpretation}}

Using the covariance interpretation of quantum metric and Berry curvature in Eq.~(\ref{gmunu_generator_interpretation}), we may write the self-bound in Eq.~(\ref{selfbound_matrix_negative_diagonal}) as
\begin{eqnarray}
\langle\Delta\Lambda_{\alpha}^{2}\rangle\geq\frac{1}{4}\langle[\Lambda_{\alpha},\Lambda_{\mu}]\rangle
\left(\frac{1}{2}\langle\left\{\Delta\Lambda,\Delta\Lambda\right\}\rangle\right)^{\mu\nu}
\langle[\Lambda_{\nu},\Lambda_{\alpha}]\rangle,
\nonumber \\
\label{selfbound_uncertainty_interpretation}
\end{eqnarray}
where $\left(\frac{1}{2}\langle\left\{\Delta\Lambda,\Delta\Lambda\right\}\rangle\right)^{\mu\nu}$ stands for the inverse matrix of the covariance matrix,  rendering a single-observable uncertainty relation. Further using the adjugate matrix in Eq.~(\ref{ginverse_adjgdetg}), one can also write it as
\begin{eqnarray}
&&4\det\left(\frac{1}{2}\langle\left\{\Delta\Lambda,\Delta\Lambda\right\}\rangle\right)\times 
\langle\Delta\Lambda_{\alpha}^{2}\rangle\geq
\nonumber \\
&&\langle[\Lambda_{\alpha},\Lambda_{\mu}]\rangle\,
{\rm adj}\left(\frac{1}{2}\langle\left\{\Delta\Lambda,\Delta\Lambda\right\}\rangle\right)^{\mu\nu}
\langle[\Lambda_{\nu},\Lambda_{\alpha}]\rangle,
\label{selfbound_uncertainty_interpretation_adjugate_mat}
\end{eqnarray}
which is an inequality between commutators and covariances of the operators. In this uncertainty interpretation, what precisely is the parameter space of the generators $\Lambda_{\mu}$ is unimportant. Because any operator can be regarded as a generator in some space, one arrives at an uncertainty relation between any set of operators $\Lambda_{\mu}$.

In comparison with many known multi-observable uncertainty relations in the literature\cite{Robertson34,Synge71,Maccone14,Chen15,Qin16,Song16,Song17,Chen17,Dodonov18,Dodonov25}, one sees that Eq.~(\ref{selfbound_uncertainty_interpretation}) is a bound for the variance of each individual operator $\langle\Delta\Lambda_{\alpha}^{2}\rangle$. This bound can then be multiplied $\prod_{\mu}\langle\Delta\Lambda_{\alpha}^{2}\rangle$ or added together $\sum_{\mu}\langle\Delta\Lambda_{\alpha}^{2}\rangle$ to construct composite bounds for multiple operators, and compare with those in the literature to see whether our bound is the tighter one, which is an issue that should be addressed elsewhere. We also remark that the interplay between QCRB, quantum geometry, multi-parameter estimation, incompatibility, and uncertainty relations in quantum measurements has been investigated intensively, such as in the Rabi oscillation measurement on nitrogen-vacancy centers\cite{Li24}. In comparison with these previous works, the significance of the present work is to point out that the QCRB itself already implies an uncertainty relation, and moreover it is an uncertainty relation for each individual observable. In Appendix \ref{apx:significance_uncertainty_individual}, we further elaborate that our single-observable uncertainty relation cannot be extracted from those of multiple observables by naive subtraction or division, pointing to the significance of our formalism.

For the two-operator cases $\mu=\left\{x,y\right\}$, using the inverse of a $2\times 2$ matrix, Eq.~(\ref{selfbound_uncertainty_interpretation}) yields 
\begin{eqnarray}
&&\langle\Delta\Lambda_{x}^{2}\rangle\langle\Delta\Lambda_{y}^{2}\rangle\geq
\frac{1}{4}|\langle\left\{\Delta\Lambda_{x},\Delta\Lambda_{y}\right\}\rangle|^{2}
+\frac{1}{4}|\langle\left[\Lambda_{x},\Lambda_{y}\right]\rangle|^{2},
\nonumber \\
\label{Robertson_Schrodinger_Heisenberg}
\end{eqnarray}
rendering the familiar Robertson-Schr\"{o}dinger uncertainty relation between any two operators $\Lambda_{x}$ and $\Lambda_{y}$\cite{Robertson29,Schrodinger30}. Thus the Robertson-Schr\"{o}dinger uncertainty relation can be thought of as a consequence of the Belavkin version of QCRB. The generalization to larger number of operators is straightforward but quickly becomes very lengthy, so we will limit our discussion to three operators $\left\{\Lambda_{x},\Lambda_{y},\Lambda_{z}\right\}$ formulated explicitly in Appendix \ref{apx:index_explicity_inequalities_3D} that are already relevant to many physical systems, as will be demonstrated in Sec.~\ref{sec:constraints_quantum_geometry_solids}.


\section{Constraints on quantum geometric properties of solids \label{sec:constraints_quantum_geometry_solids}}

\subsection{Quantum geometric constraints in solids}

In this section, we elaborate the nontrivial consequences of the bound on quantum metric revealed in Sec.~\ref{sec:geometric_interpretation} for solids. For semiconductors, insulators, or superconductors that have $N_{-}$ valence or quasihole bands, the notion of quantum geometry arises when one considers the fully antisymmetric filled band Bloch state that treats the crystalline momentum ${\bf k}=(k^{x},k^{y}...k^{D})$ as the tuning parameter
\begin{eqnarray}
|\psi({\bf k})\rangle\equiv \frac{1}{\sqrt{ N_-!}}\epsilon^{a_1\cdots a_{N_-}}|n_{a_1}\rangle|n_{a_2}\rangle\cdots|n_{a_{N_{-}}}\rangle.
\label{eq:fermi}
\end{eqnarray}
whose quantum metric and Berry curvature are\cite{Matsuura10,vonGersdorff21_metric_curvature}
\begin{eqnarray}
&&g_{\mu\nu}=\frac{1}{2}\sum_{nm}\langle\partial_{\mu}n|m\rangle\langle m|\partial_{\nu}n\rangle+(\mu\leftrightarrow \nu),
\nonumber \\
&&\Omega_{\mu\nu}=i\sum_{nm}\langle\partial_{\mu}n|m\rangle\langle m|\partial_{\nu}n\rangle-(\mu\leftrightarrow \nu),
\label{gmunu_Omegamunu_nm_expression}
\end{eqnarray}
where $|n\rangle$ and $|m\rangle$ denote the valence (or quasihole) and conduction (or quasiparticle) band states, respectively. For the sake of our discussion, we write the bound on the diagonal elements of quantum metric schematically as 
\begin{eqnarray}
g_{\mu\mu}({\bf k})\geq b_{\mu\mu}({\bf k}),
\label{gmumu_Bmumu}
\end{eqnarray}
where $b_{\mu\mu}$ is the right hand side of Eq.~(\ref{QCRB_pure_state_2D}) for 2D solids, and the right hand side of Eq.~(\ref{selfbound_3D_index_explicit}) for 3D solids.

We now demonstrate that the each direction of the gauge-invariant part of the spread of valence or quasihole band Wannier function is bounded due to Eq.~(\ref{gmumu_Bmumu}). The spread of Wannier function is defined by\cite{Marzari97,Marzari12,Porlles23_singlet_SC_metric}
\begin{eqnarray}
&&\Omega=\sum_{n}\left[\langle{\bf 0}n|r^{2}|{\bf 0}n\rangle-\langle{\bf 0}n|{\bf r}|{\bf 0}n\rangle^{2}\right]=\Omega_{I}+\tilde{\Omega},
\label{Omega_original}
\end{eqnarray}
that can be separated to the gauge-invariant $\Omega_{I}$ and gauge-dependent $\tilde{\Omega}$, where $|{\bf R}n\rangle$ is the Wannier state for valence or quasihole band $n$ centering at homecell ${\bf R}$, and ${\bf r}$ is the position operator. Introducing the fidelity number calculated from the momentum integration of the quantum metric\cite{deSousa23_fidelity_marker}
\begin{eqnarray}
&&{\cal G}_{\mu\nu}=\int\frac{d^{D}{\bf k}}{(2\pi)^{D}}g_{\mu\nu}({\bf k}),
\label{Gmunu_definition}
\end{eqnarray} 
and decomposing the spread into each spatial direction
\begin{eqnarray}
\Omega_{I}=\sum_{\mu=1}^{D}\Omega_{I,\mu},
\end{eqnarray}
it is known that each direction of the gauge-invariant part of the spread is given by the diagonal elements of the fidelity number\cite{Marzari97,Marzari12,Souza08}, and hence bounded by
\begin{eqnarray}
&&\Omega_{I,\mu}
=\frac{V_{\rm cell}}{\hbar^{D-2}}{\cal G}_{\mu\mu}\geq\frac{V_{\rm cell}}{\hbar^{D-2}}\int\frac{d^{D}{\bf k}}{(2\pi)^{D}}b_{\mu\mu}({\bf k}),
\label{OmegaImu_Gmumu_bound_general}
\end{eqnarray}
owing to Eq.~(\ref{gmumu_Bmumu}), where $V_{\rm cell}$ is the volume of the unit cell. Thus the integration of Eq.~(\ref{gmumu_Bmumu}) over the momentum space sets a limit on the spread $\Omega_{I,\mu}$ in each spatial direction.

Equation (\ref{OmegaImu_Gmumu_bound_general}) also has direct consequences on optical sum rules of 2D and 3D semiconductors. For 2D semiconductors, denoting $O_{\mu}(\omega)$ as the absorbance under light with frequency $\omega$ and polarization ${\hat{\boldsymbol\mu}}$, a relation between $\Omega_{I,\mu}$ and the frequency integration of $O_{\mu}(\omega)/\omega$ 
\begin{eqnarray}
&&\Omega_{I,\mu}=A_{\rm cell}\int_{0}^{\infty}d\omega\,\frac{1}{4\pi\omega}\left[\frac{{\cal O}_{\mu}(\omega)}{\pi\alpha}\right]
\nonumber \\
&&\geq A_{\rm cell}\int\frac{d^{2}{\bf k}}{(2\pi)^{2}}b_{\mu\mu}({\bf k}),
\label{OmegaImu_opacity_bound}
\end{eqnarray}
dictates that the frequency integration is bounded by the momentum integration of Eq.~(\ref{gmumu_Bmumu}), where $A_{\rm cell}$ is the area of the unit cell, and $\alpha\approx 1/137$ is the fine-structure constant\cite{CardenasCastillo24_spread_Wannier}. For 3D semiconductors, consider the frequency integration of the imaginary part of dielectric function $\varepsilon_{2,\mu}(\omega)$ under light with frequency $\omega$ and polarization ${\hat{\boldsymbol\mu}}$, which is related to the spread by\cite{CardenasCastillo24_spread_Wannier}
\begin{eqnarray}
&&\Omega_{I,\mu}=\frac{V_{\rm cell}\varepsilon_{0}}{\pi e^{2}}\int_{0}^{\infty}d(\hbar\omega)\,\varepsilon_{2,\mu}(\omega)
\nonumber \\
&&\geq\frac{V_{\rm cell}}{\hbar}\int\frac{d^{3}{\bf k}}{(2\pi)^{3}}b_{\mu\mu}({\bf k}),
\end{eqnarray}
and hence the frequency integration of $\varepsilon_{2,\mu}(\omega)$ in 3D is also bounded by the momentum integration of Eq.~(\ref{gmumu_Bmumu}). We are lead to conclude that the bound in Eq.~(\ref{gmumu_Bmumu}) highly constrains the optical sum rules of semiconductors, especially in materials that have nonzero Berry curvature.

\subsection{Geometric interpretation in 3D class AII topological insulators}

To demonstrate the geometric bounds in Sec.~\ref{sec:geometric_interpretation} for 3D parameter spaces, we resort to 3D time-reversal symmetric TIs in class AII, where the crystalline momentum ${\bf k}=(k^{x},k^{y},k^{z})$ naturally provides a 3D parameter space $\mu=(x,y,z)$. This example is motivated by a recent surge of interest in the momentum space quantum metric of semiconductors and insulators, which has been shown to be responsible for many dielectric and optical properties\cite{Komissarov24,Chen25_optical_marker,Porlles26_semiconductor_metric}, as well as the rich quantum geometric properties in 3D topological materials like vacuum Einstein equation and constant Ricci scalar\cite{Matsuura10,Chen25_genetic_TI_TSC}, and the link to the topological order\cite{vonGersdorff21_metric_curvature,Mera22}. The pristine model relevant to real materials like Bi$_{2}$Se$_{3}$ and Bi$_{2}$Te$_{3}$ is described by a $4\times 4$ Dirac Hamiltonian in momentum space $H({\bf k})={\bf d(k)}\cdot{\boldsymbol\Gamma}$, with $\Gamma_{i}$ the Dirac matrices\cite{Zhang09,Liu10}
\begin{eqnarray}
\Gamma_{1\sim 5}=\left\{\sigma^{x}\otimes\tau^{x},\sigma^{y}\otimes\tau^{x},\sigma^{z}\otimes\tau^{x},
I_{\sigma}\otimes\tau^{y},I_{\sigma}\otimes\tau^{z}\right\},
\nonumber \\
\end{eqnarray} 
where the basis is $\left(c_{{\bf k}s\uparrow},c_{{\bf k}p\uparrow},c_{{\bf k}s\downarrow},c_{{\bf k}p\downarrow}\right)^{T}$, with $s$ and $p$ stand for the $P1_{-}^{+}$ and $P2_{+}^{-}$ orbitals in real materials. The ${\bf d(k)}$-vector that describes the momentum dependence has the components\cite{Chen20_absence_edge_current}
\begin{eqnarray}
&&d_{5}=M+6B-2B\sum_{i=1}^{3}\cos \left(k_{i}a/\hbar\right),
\nonumber \\
&&d_{1}=A\sin \left(k_{y}a/\hbar\right),\;\;\;
d_{2}=-A\sin \left(k_{x}a/\hbar\right),
\nonumber \\
&&d_{3}=0,\;\;\;
d_{4}=A\sin \left(k_{z}a/\hbar\right).
\label{3D_class_AII_kspace_model}
\end{eqnarray}
This pristine model does not serve our purpose because it possesses no Berry curvature owing to time-reversal and inversion symmetries\cite{Xiao10}, and hence the self-bound on quantum metric in Eq.~(\ref{selfbound_matrix_negative_diagonal}) becomes trivially $g_{\mu\mu}\geq 0$. To make the self-bound nontrivial, we add a hypothetical magnetic field into the game to break the time-reversal symmetry, leading to the Hamiltonian 
\begin{eqnarray}
H({\bf k})={\bf d(k)}\cdot{\boldsymbol\Gamma}+{\bf B}\cdot{\boldsymbol\sigma},
\label{Hamiltonian_3D_AII_TI_Bfield}
\end{eqnarray}
where ${\boldsymbol\sigma}$ is the spin operator. The presence of a nonzero magnetic field ${\bf B}=(B_{x},B_{y},B_{z})$ pointing at arbitrary direction generates a nonzero Berry curvature, hence the right hand side of Eq.~(\ref{selfbound_matrix_negative_diagonal}) becomes nonzero. For concreteness, we choose the band parameters to be $M=-0.3$, $A=2.87$, $B=0.3$ in units of eV that are relevant to Bi$_{2}$Se$_{3}$ and Bi$_{2}$Te$_{3}$, while investigating different values of the magnetic field ${\bf B}$.

\begin{figure}[ht]
\begin{center}
\includegraphics[clip=true,width=0.99\columnwidth]{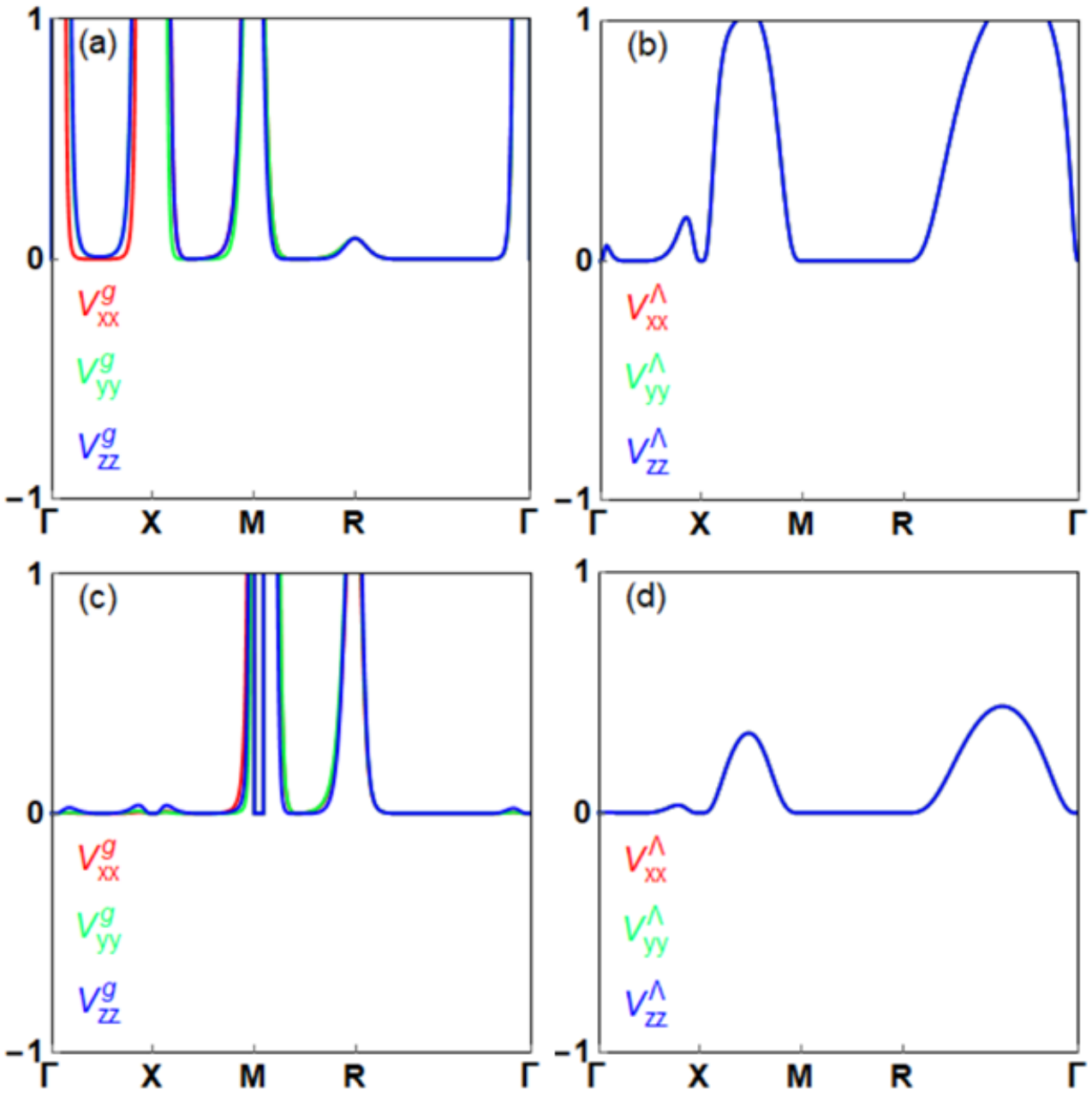}
\caption{(a) The bound on quantum metric represented by $(V_{xx}^{g},V_{yy}^{g},V_{zz}^{g})$ and (b) the uncertainty relation for spin operators represented by $(V_{xx}^{\Lambda},V_{yy}^{\Lambda},V_{zz}^{\Lambda})$ in the lattice model of 3D class AII TI in the presence of a magnetic field $(B_{x},B_{y},B_{z})=(0.1,0.2,0.3)$, plotted along the high-symmetry line $\Gamma-X-M-R-\Gamma$. (c) and (d) show the same quantities but in an extremely large magnetic field $(B_{x},B_{y},B_{z})=(0.5,1,2)$. One sees that all these quantities remain positive everywhere, indicating that the bounds and uncertainty relations are always satisfied. } 
\label{fig:QCRB_3D_AII_C_V}
\end{center}
\end{figure}

We proceed to take the two lower energy eigenstates $|n_{1}\rangle$ and $|n_{2}\rangle$ to construct the fully antisymmetric fermionic state $|\psi({\bf k})\rangle=\left(|n_{1}\rangle|n_{2}\rangle-|n_{2}\rangle|n_{1}\rangle\right)/\sqrt{2}$ according to Eq.~(\ref{eq:fermi}), 
and calculate the quantum metric and Berry curvature according to Eq.~(\ref{gmunu_Omegamunu_nm_expression}), which can then be used to verify Eq.~(\ref{selfbound_3D_index_explicit}) at a specific momentum ${\bf k}$ at some value of magnetic field ${\bf B}$. To systematically verify the bound, we define $V_{\mu\mu}^{g}$ as the left hand side of Eq.~(\ref{QCRB_gmunu_adjugate_mat_diagonal}) minus the right hand side (see Appendix \ref{apx:index_explicity_inequalities_3D} for matrix elements), which should be always positive at any momentum ${\bf k}$ regardless of the system parameters. To demonstrate this, we choose to plot $V_{\mu\mu}^{g}$ along the high symmetry lines $\Gamma-X-M-R-\Gamma$ in the Brillouin zone, with the high symmetry points located at $\Gamma=(0,0,0)$, $X=(\pi,0,0)$, $M=(\pi,\pi,0)$, and $R=(\pi,\pi,\pi)$. The result shown in Fig.~\ref{fig:QCRB_3D_AII_C_V} (a) at a small magnetic field $(B_{x},B_{y},B_{z})=(0.1,0.2,0.3)$ in units of eV, chosen for no particular reason, indeed verifies that all the three components $(V_{xx}^{g},V_{yy}^{g},V_{zz}^{g})$ are always positive along the high symmetry line, and all three are very close in magnitude. There are regions where $V_{\mu\mu}^{g}$ is numerically zero, such as in a large part of the $R-\Gamma$ section, owing to the fact that $\det g_{\mu\nu}$ becomes zero, meaning that the equality condition of Eq.~(\ref{QCRB_gmunu_adjugate_mat_diagonal}) is satisfied by zero equals to zero. Even if we change the magnetic field to be an extremely large value $(B_{x},B_{y},B_{z})=(0.5,1,2)$, again chosen for no particular reason, the $(V_{xx}^{g},V_{yy}^{g},V_{zz}^{g})$ still remain positive as shown in Fig.~\ref{fig:QCRB_3D_AII_C_V} (c), suggesting that the bound on quantum metric is always satisfied under any condition.

\subsection{Uncertainty relation for spin operators in 3D class AII topological insulators}

We proceed to demonstrate the uncertainty interpretation in Sec.~\ref{sec:uncertainty_interpretation} for the same model of 3D TI under magnetic field. For concreteness, we choose the operators to be the three spin operators $\left\{\Lambda_{x},\Lambda_{y},\Lambda_{z}\right\}=\left\{\sigma_{x},\sigma_{y},\sigma_{z}\right\}$ that couple to the magnetic field according to Eq.~(\ref{Hamiltonian_3D_AII_TI_Bfield}), and examine the corresponding uncertainty relations for $\langle\Delta\sigma_{\mu}^{2}\rangle$. In particular, because the spin operator is a single-particle operator, we choose to examine the uncertainty relation for the lowest eigenenergy state
\begin{eqnarray}
|\psi({\bf k})\rangle=|n_{1}({\bf k})\rangle,
\end{eqnarray}
and calculate the expectation values by $\langle{\hat O}\rangle=\langle n_{1}|{\hat O}|n_{1}\rangle$ for any operator ${\hat O}$ that appears in Eq.~(\ref{selfbound_uncertainty_interpretation_adjugate_mat}).

In the pristine model that has no magnetic field ${\bf B}={\bf 0}$, time-reversal symmetry ensures that the spin expectation value is zero, so Eq.~(\ref{selfbound_uncertainty_interpretation_adjugate_mat}) becomes trivially $\langle\Delta\sigma_{\mu}^{2}\rangle\geq 0$. It is the presence of magnetic field ${\bf B}\neq {\bf 0}$ that causes nonzero spin expectation values and makes Eq.~(\ref{selfbound_uncertainty_interpretation_adjugate_mat}) nontrivial. To quantify this magnetic field effect, we define $V_{\mu\mu}^{\Lambda}$ as the left hand side of Eq.~(\ref{selfbound_uncertainty_interpretation_adjugate_mat}) minus the right hand side (explicit matrix elements involved are given in Appendix \ref{apx:index_explicity_inequalities_3D}), which should always be positive. The numerical results in Fig.~\ref{fig:QCRB_3D_AII_C_V} (b) for the small magnetic field $(B_{x},B_{y},B_{z})=(0.1,0.2,0.3)$ and in Fig.~\ref{fig:QCRB_3D_AII_C_V} (d) for the large magnetic field $(B_{x},B_{y},B_{z})=(0.5,1,2)$ indeed confirm the validity of Eq.~(\ref{selfbound_uncertainty_interpretation}), where one sees that the three components $(V_{xx}^{\Lambda},V_{yy}^{\Lambda},V_{zz}^{\Lambda})$ are numerically identical, and are always positive at any momentum regardless of the strength of the magnetic field. Note that $V_{\mu\mu}^{\Lambda}=0$ in the $M-R$ section of the high-symmetry line, indicating that the uncertainty relation is exactly at the low bound, which is caused by the vanishing determinant $\det g_{\mu\nu}=0$ of the covariance matrix in Eq.~(\ref{detgmunu_gxxgyygzz}), causing the equality condition to be satisfied by zero equals to zero.


\section{Conclusions}

In summary, we elaborate the nontrivial consequences of the Belavkin version of QCRB in the areas of quantum geometry and uncertainty relation. In the geometric interpretation, the Belavkin version of QCRB manifests as a geometric bound on the diagonal elements of quantum metric caused by the quantum metric itself and the Berry curvature, rendering an inequality that is applicable to parameter spaces of any dimension. This bound is particularly nontrivial in systems that have nonzero Berry curvature, as demonstrated for a theoretical model of 3D TI in the presence of a magnetic field. The bound is further shown to constrain the spread of Wannier function and optical sum rules of semiconductors in each spatial direction, such as the absorbance in 2D and dielectric function in 3D, as long as Berry curvature is nonzero. In the uncertainty interpretation, the Belavkin version of QCRB yields a single-observable uncertainty relation for the variance of each individual operator caused by the covariances and commutators of all the operators in the system, as demonstrated by the three spin operators in the 3D TI. The two-operator version coincides with and hence demonstrates the QCRB interpretation of the Robertson-Schr\"{o}dinger uncertainty relation. As these bounds on quantum metric and single-observable uncertainty relations are not limited to a specific parameter space or a particular set of operators, we anticipate that their applications are abundant, which await to be further explored.

\begin{acknowledgments}

We acknowledge the support of the INCT project Advanced Quantum Materials, 
involving the Brazilian agencies CNPq (Proc. 408766/2024-7), FAPESP, and CAPES, as well as the financial support of the productivity in research fellowship from CNPq. 

\end{acknowledgments}

\appendix

\section{Index-explicit inequalities in 3D \label{apx:index_explicity_inequalities_3D}}

For 3D parameter spaces, the inverse of quantum metric takes the form
\begin{eqnarray}
&&g^{\mu\nu}=\frac{1}{\det g_{\mu\nu}}\times
\nonumber \\
&&\left(\begin{array}{ccc}
g_{yy}g_{zz}-g_{yz}^{2} & g_{xz}g_{yz}-g_{xy}g_{zz} & g_{xy}g_{yz}-g_{xz}g_{yy} \\
g_{xz}g_{yz}-g_{xy}g_{zz} & g_{xx}g_{zz}-g_{xz}^{2} & g_{xy}g_{xz}-g_{xx}g_{yz}
\\
g_{xy}g_{yz}-g_{xz}g_{yy} & g_{xy}g_{xz}-g_{xx}g_{yz} & g_{xx}g_{yy}-g_{xy}^{2}
\end{array}\right).
\nonumber \\
\label{gmunu_inverse_3D}
\end{eqnarray}
where the determinant of the quantum metric is 
\begin{eqnarray}
\det g_{\mu\nu}
&=&g_{xx}g_{yy}g_{zz}+2g_{xy}g_{yz}g_{zx}
\nonumber \\
&-&g_{xx}g_{yz}^{2}-g_{yy}g_{zx}^{2}-g_{zz}g_{xy}^{2}.
\label{detgmunu_gxxgyygzz}
\end{eqnarray}
Using Eqs.~(\ref{gmunu_inverse_3D}) and (\ref{detgmunu_gxxgyygzz}), the inequalities in Eq.~(\ref{QCRB_pure_state_3D}) can be completely expressed in terms of lower indices $g_{\mu\nu}$ and $\Omega_{\mu\nu}$, rendering
\begin{widetext}
\begin{eqnarray}
&&g_{xx}\geq\frac{\Omega_{xy}^{2}(g_{zz}g_{xx}-g_{zx}^{2})+\Omega_{zx}^{2}(g_{xx}g_{yy}-g_{xy}^{2})
-2\Omega_{xy}\Omega_{zx}(g_{xy}g_{zx}-g_{xx}g_{yz})}{4g_{xx}g_{yy}g_{zz}+8g_{xy}g_{yz}g_{zx}
-4g_{xx}g_{yz}^{2}-4g_{yy}g_{zx}^{2}-4g_{zz}g_{xy}^{2}}\geq 0,
\nonumber \\
&&g_{yy}\geq\frac{\Omega_{yz}^{2}(g_{xx}g_{yy}-g_{xy}^{2})+\Omega_{xy}^{2}(g_{yy}g_{zz}-g_{yz}^{2})
-2\Omega_{yz}\Omega_{xy}(g_{yz}g_{xy}-g_{yy}g_{zx})}{4g_{xx}g_{yy}g_{zz}+8g_{xy}g_{yz}g_{zx}
-4g_{xx}g_{yz}^{2}-4g_{yy}g_{zx}^{2}-4g_{zz}g_{xy}^{2}}\geq 0,
\nonumber \\
&&g_{zz}\geq\frac{\Omega_{zx}^{2}(g_{yy}g_{zz}-g_{yz}^{2})+\Omega_{yz}^{2}(g_{zz}g_{xx}-g_{zx}^{2})
-2\Omega_{zx}\Omega_{yz}(g_{zx}g_{yz}-g_{zz}g_{xy})}{4g_{xx}g_{yy}g_{zz}+8g_{xy}g_{yz}g_{zx}
-4g_{xx}g_{yz}^{2}-4g_{yy}g_{zx}^{2}-4g_{zz}g_{xy}^{2}}\geq 0,
\label{selfbound_3D_index_explicit}
\end{eqnarray}
which serve as constraints on $g_{\mu\nu}$ and $\Omega_{\mu\nu}$ in 3D parameter spaces. 


To implement uncertainty relations, we first write the determinant in Eq.~(\ref{detgmunu_gxxgyygzz}) in terms of the covariances
\begin{eqnarray}
\det g_{\mu\nu}
&=&\langle\Delta\Lambda_{x}^{2}\rangle\langle\Delta\Lambda_{y}^{2}\rangle\langle\Delta\Lambda_{z}^{2}\rangle
+\frac{1}{4}\langle\left\{\Delta\Lambda_{x},\Delta\Lambda_{y}\right\}\rangle
\langle\left\{\Delta\Lambda_{y},\Delta\Lambda_{z}\right\}\rangle
\langle\left\{\Delta\Lambda_{z},\Delta\Lambda_{x}\right\}\rangle
\nonumber \\
&-&\frac{1}{4}\langle\left\{\Delta\Lambda_{y},\Delta\Lambda_{z}\right\}\rangle^{2}\langle\Delta\Lambda_{x}^{2}\rangle
-\frac{1}{4}\langle\left\{\Delta\Lambda_{z},\Delta\Lambda_{x}\right\}\rangle^{2}\langle\Delta\Lambda_{y}^{2}\rangle
-\frac{1}{4}\langle\left\{\Delta\Lambda_{x},\Delta\Lambda_{y}\right\}\rangle^{2}\langle\Delta\Lambda_{z}^{2}\rangle.
\label{detg_3D_uncertainty_interpretation}
\end{eqnarray}
Using this and the generator interpretation in Eq.~(\ref{gmunu_generator_interpretation}), the inequalities in Eq.~(\ref{selfbound_3D_index_explicit}) become
\begin{eqnarray}
&&\langle\Delta\Lambda_{x}^{2}\rangle\geq\frac{1}{4\det g_{\mu\nu}}\left(-\langle[\Lambda_{x},\Lambda_{y}]\rangle^{2}\langle\Delta\Lambda_{z}^{2}\rangle\langle\Delta\Lambda_{x}^{2}\rangle
+\frac{1}{4}\langle[\Lambda_{x},\Lambda_{y}]\rangle^{2}\langle\left\{\Delta\Lambda_{z},\Delta\Lambda_{x}\right\}\rangle^{2}\right.
\nonumber \\
&&+\frac{1}{2}\langle[\Lambda_{x},\Lambda_{y}]\rangle\langle[\Lambda_{z},\Lambda_{x}]\rangle
\langle\left\{\Delta\Lambda_{x},\Delta\Lambda_{y}\right\}\rangle\langle\left\{\Delta\Lambda_{z},\Delta\Lambda_{x}\right\}\rangle
-\langle[\Lambda_{x},\Lambda_{y}]\rangle\langle[\Lambda_{z},\Lambda_{x}]\rangle\langle\left\{\Delta\Lambda_{y},\Delta\Lambda_{z}\right\}\rangle
\langle\Delta\Lambda_{x}^{2}\rangle
\nonumber \\
&&\left.-\langle[\Lambda_{z},\Lambda_{x}]\rangle^{2}\langle\Delta\Lambda_{x}^{2}\rangle\langle\Delta\Lambda_{y}^{2}\rangle
+\frac{1}{4}\langle[\Lambda_{z},\Lambda_{x}]\rangle^{2}\langle\left\{\Delta\Lambda_{x},\Delta\Lambda_{y}\right\}\rangle^{2}\right)
\geq 0,
\nonumber \\
&&\langle\Delta\Lambda_{y}^{2}\rangle\geq\frac{1}{4\det g_{\mu\nu}}\left(-\langle[\Lambda_{y},\Lambda_{z}]\rangle^{2}\langle\Delta\Lambda_{x}^{2}\rangle\langle\Delta\Lambda_{y}^{2}\rangle
+\frac{1}{4}\langle[\Lambda_{y},\Lambda_{z}]\rangle^{2}\langle\left\{\Delta\Lambda_{x},\Delta\Lambda_{y}\right\}\rangle^{2}\right.
\nonumber \\
&&+\frac{1}{2}\langle[\Lambda_{y},\Lambda_{z}]\rangle\langle[\Lambda_{x},\Lambda_{y}]\rangle
\langle\left\{\Delta\Lambda_{y},\Delta\Lambda_{z}\right\}\rangle\langle\left\{\Delta\Lambda_{x},\Delta\Lambda_{y}\right\}\rangle
-\langle[\Lambda_{y},\Lambda_{z}]\rangle\langle[\Lambda_{x},\Lambda_{y}]\rangle\langle\left\{\Delta\Lambda_{z},\Delta\Lambda_{x}\right\}\rangle
\langle\Delta\Lambda_{y}^{2}\rangle
\nonumber \\
&&\left.-\langle[\Lambda_{x},\Lambda_{y}]\rangle^{2}\langle\Delta\Lambda_{y}^{2}\rangle\langle\Delta\Lambda_{z}^{2}\rangle
+\frac{1}{4}\langle[\Lambda_{x},\Lambda_{y}]\rangle^{2}\langle\left\{\Delta\Lambda_{y},\Delta\Lambda_{z}\right\}\rangle^{2}\right)
\geq 0,
\nonumber \\
&&\langle\Delta\Lambda_{z}^{2}\rangle\geq\frac{1}{4\det g_{\mu\nu}}\left(-\langle[\Lambda_{z},\Lambda_{x}]\rangle^{2}\langle\Delta\Lambda_{y}^{2}\rangle\langle\Delta\Lambda_{z}^{2}\rangle
+\frac{1}{4}\langle[\Lambda_{z},\Lambda_{x}]\rangle^{2}\langle\left\{\Delta\Lambda_{y},\Delta\Lambda_{z}\right\}\rangle^{2}\right.
\nonumber \\
&&+\frac{1}{2}\langle[\Lambda_{z},\Lambda_{x}]\rangle\langle[\Lambda_{y},\Lambda_{z}]\rangle
\langle\left\{\Delta\Lambda_{z},\Delta\Lambda_{x}\right\}\rangle\langle\left\{\Delta\Lambda_{y},\Delta\Lambda_{z}\right\}\rangle
-\langle[\Lambda_{z},\Lambda_{x}]\rangle\langle[\Lambda_{y},\Lambda_{z}]\rangle\langle\left\{\Delta\Lambda_{x},\Delta\Lambda_{y}\right\}\rangle
\langle\Delta\Lambda_{z}^{2}\rangle
\nonumber \\
&&\left.-\langle[\Lambda_{y},\Lambda_{z}]\rangle^{2}\langle\Delta\Lambda_{z}^{2}\rangle\langle\Delta\Lambda_{x}^{2}\rangle
+\frac{1}{4}\langle[\Lambda_{y},\Lambda_{z}]\rangle^{2}\langle\left\{\Delta\Lambda_{z},\Delta\Lambda_{x}\right\}\rangle^{2}\right)
\geq 0.
\label{QCRB_pure_state_3D_uncertainty_interpretation}
\end{eqnarray}
\end{widetext}
These three-operator uncertainty relations are major results of the present work, which constrain the variance of an operator $\langle\Delta\Lambda_{\mu}^{2}\rangle$ by the different kinds of covariances of all the three operators in the problem.

It should be emphasized again that this self-bound is meaningful only if the determinant of the quantum metric $\det g_{\mu\nu}$ or equivalently the determinant of the covariance matrix $\det\frac{1}{2}\langle\left\{\Delta\Lambda_{\mu},\Delta\Lambda_{\nu}\right\}\rangle$ is nonzero according to Eq.~(\ref{ginverse_adjgdetg}). Two counter examples are given below. The first are the angular momentum operators $\Lambda_{\mu}=\left\{L_{x},L_{y},L_{z}\right\}$ applied to atomic orbitals $|\psi\rangle=|\ell,m\rangle$, whose covariances can be calculated from
\begin{eqnarray}
&&\langle L_{x}\rangle=\langle L_{y}\rangle=0,\;\;\;\langle L_{z}\rangle=m\hbar,
\;\;\;\langle L_{z}^{2}\rangle=m^{2}\hbar^{2}
\nonumber \\
&&\langle L_{x}^{2}\rangle=\langle L_{y}^{2}\rangle=\frac{\hbar^{2}}{2}\left(\ell^{2}+\ell-m^{2}\right),
\end{eqnarray}
yielding a $3\times 3$ covariance matrix that has zero determinant $\det\frac{1}{2}\langle\left\{\Delta L_{\mu},\Delta L_{\nu}\right\}\rangle=0$,
and consequently the uncertainty relations in Eq.~(\ref{QCRB_pure_state_3D_uncertainty_interpretation}) become trivially zero equals to zero if we multiply both sides by the determinant. The second example concerns a generic $2\times 2$ Dirac Hamiltonian expanded by Pauli matrices 
\begin{eqnarray}
H={\bf d}\cdot{\boldsymbol\sigma}=d_{1}\sigma_{1}+d_{2}\sigma_{2}+d_{3}\sigma_{3}\;,
\end{eqnarray}
and we choose $|\psi\rangle$ to be the lower energy state with eigenenergy $-d=-\sqrt{d_{1}^{2}+d_{2}^{2}+d_{3}^{2}}$. In this case, if we treat the Pauli matrices as the generators $\Lambda_{\mu}=\left\{\sigma_{1},\sigma_{2},\sigma_{3}\right\}$, then the covariances of the generators are 
\begin{eqnarray}
&&\langle\sigma_{\mu}\rangle=-\frac{d_{\mu}}{d},\;\;\;
\langle[\sigma_{\mu},\sigma_{\nu}]\rangle=-2i\epsilon^{\mu\nu\lambda}\frac{d_{\lambda}}{d},
\nonumber \\
&&\langle\left\{\Delta\sigma_{\mu},\Delta\sigma_{\nu}\right\}\rangle=-2\frac{d_{\mu}d_{\nu}}{d^{2}},
\end{eqnarray}
leading to a vanishing determinant $\det\frac{1}{2}\langle\left\{\Delta\sigma_{\mu},\Delta\sigma_{\nu}\right\}\rangle=0$ of the $3\times 3$ covariance matrix, and hence the uncertainty relations in Eq.~(\ref{QCRB_pure_state_3D_uncertainty_interpretation}) for the Pauli matrices are also trivially zero equals to zero if both sides are multiplied by the determinant. 



\section{Significance of the single-observable uncertainty relation \label{apx:significance_uncertainty_individual}}

Although many multi-observable uncertainty relations have been proposed since the early development of quantum mechanics, they do not automatically imply any single-observable uncertainty relation. To demonstrate this statement, suppose we take any of the three-operator uncertainty relations that has been proposed and compared with the two-operator versions like Heisenberg or Robertson-Schr\"{o}dinger, written schematically as 
\begin{eqnarray}
&&\langle\Delta\Lambda_{x}^{2}\rangle\langle\Delta\Lambda_{y}^{2}\rangle\langle\Delta\Lambda_{z}^{2}\rangle\geq C_{3},
\nonumber \\
&&\langle\Delta\Lambda_{y}^{2}\rangle\langle\Delta\Lambda_{z}^{2}\rangle\geq C_{2},
\end{eqnarray}
where $C_{3}$ and $C_{2}$ consist of covariances and commutators of all the operators in the problem. Then it is tempted to divide the two inequalities to claim the uncertainty relation for a single observable $\langle\Delta\Lambda_{x}^{2}\rangle\geq C_{3}/C_{2}$, but this is not always valid. For instance, $7>6$ and $5>3$ do not imply $7/5$ is greater than $6/3$.

Similarly, several uncertainty relations for the sum of multiple variances have also been proposed, but they also do not imply an uncertainty relation for individual observables. As an example, consider two inequalities containing the sum of three and two variances 
\begin{eqnarray}
&&\langle\Delta\Lambda_{x}^{2}\rangle+\langle\Delta\Lambda_{y}^{2}\rangle+\langle\Delta\Lambda_{z}^{2}\rangle\geq D_{3},
\nonumber \\
&&\langle\Delta\Lambda_{y}^{2}\rangle+\langle\Delta\Lambda_{z}^{2}\rangle\geq D_{2}.
\end{eqnarray}
One may subtract the two inequalities to claim that $\langle\Delta\Lambda_{x}^{2}\rangle\geq D_{3}-D_{2}$, but again this is not always true. For instance, the examples $7>6$ and $5>3$ mentioned above do not imply $7-5$ is greater than $6-3$. From this discussion, we see that it is in general not possible to extract uncertainty relations for individual observables from inequalities containing the product, sum, or any other functions of multiple variances. Thus our formalism based on the operator version of QCRB points to an important methodology that can address variances of individual operators.

\bibliography{Literatur}

\end{document}